 \documentclass[final,5p,times,twocolumn]{elsarticle}

\usepackage{amssymb}
\usepackage{lipsum}
\usepackage{graphicx}%
\usepackage{multirow}%
\usepackage{amsmath,amssymb,amsfonts}%
\usepackage{amsthm}%
\usepackage{mathrsfs}%
\usepackage[title]{appendix}%
\usepackage{xcolor}%
\usepackage{textcomp}%
\usepackage{manyfoot}%
\usepackage{booktabs}%
\usepackage{algorithm}%
\usepackage{algorithmicx}%
\usepackage{algpseudocode}%
\usepackage{listings}%
\usepackage{orcidlink}%
\usepackage{dcolumn}%

\journal{Physics Letters B}

\begin{document}

\begin{frontmatter}

\title{The Statistical Mechanics of Hawking Radiation}

\author[first]{Noah M. MacKay\,\orcidlink{0000-0001-6625-2321}}
\affiliation[first]{organization={Universität Potsdam, Institut für Physik und Astronomie},
            addressline={Karl-Liebknecht-Straße 24/25}, 
           postcode={14476},  
           city={Potsdam},
            country={Germany}}

\begin{abstract}
Hawking radiation and black hole thermodynamics are well understood in the frameworks of quantum field theory and general relativity, with contemporary extensions in string theory, AdS/CFT, and loop quantum gravity. However, an open question remains: \textit{Can Hawking radiation be consistently described using statistical mechanics?} This challenge arises due to the ambiguous quantum nature of Hawking particles and constraints imposed by information conservation. This study develops a heuristic statistical model using a worldine Lagrangian formalism, treating Hawking particles as timelike with an effective mass and following an arbitrary statistical distribution. The flow of information is modeled as a transfer of microstates from the event horizon to the radiation background within a closed ensemble. From this framework, the Hawking particle mass $m_H=\hbar/(2\pi r_S)$ is recovered, and the worldline Lagrangian yields the black hole's thermal energy, $E=\hbar/(8\pi GM)$. Beyond preserving conventional insights, reinforcing the statistical nature of black hole evaporation, this approach introduces novel perspectives: modeling the black hole as a structured energy well containing microstates, quantifying the evaporation rate of individual microstates, and determining the total number of microstates from the horizon's surface area. This study leaves the door open for a stochastic mechanical analysis, which is introduced in the Discussion section.
\end{abstract}

\begin{keyword}
Hawking radiation \sep Black hole thermodynamics \sep Statistical mechanics 



\end{keyword}

\end{frontmatter}




\section{Introduction} \label{intro}

According to the cosmic censorship conjecture, a spherical body with radius $r<r_S$ cannot exist in equilibrium, where $r_S=2GM$ ($c=1$) is the Schwarzschild radius \cite{Schwarzschild:1916uq, Schwarzschild:1916ae}. Such a body must undergo gravitational collapse into a central singularity enclosed by an event horizon, and the distance between the singularity and the horizon is the radius $r_S$ \cite{Penrose:1964wq}. This unstable, out-of-equilibrium entity is a black hole. If the black hole is stationary, it is predicted to emit a form of blackbody radiation, leading to its gradual evaporation \cite{Hawking:1974rv, Hawking:1975vcx}; this is Hawking radiation.

Hawking radiation arises from the splitting of vacuum particle-antiparticle pairs along the event horizon. One particle escapes as a relativistic particle of mass $m_H$, while the other is captured in the black hole, effectively forming a thermodynamic system. Assuming the radiation and the black hole are in thermal equilibrium, they share a common temperature known as the Hawking temperature:
\begin{equation}\label{hawktemp}
T_{H}=\frac{\hbar}{8\pi GMk_B}.
\end{equation}
Here, the black hole's internal temperature is inversely proportional to its mass. 

It is a logical question whether black hole thermodynamics and Hawking radiation can be approached using statistical mechanics. What adds complexity to this question is the ambiguity of Hawking particles, from their species type (i.e., their distribution type) to their inherent quantum characteristics, such as spin and degeneracy factors. For instance, it is proposed that the total power output Hawking radiation consists of 81\% neutrinos, 17\% photons, and 2\% gravitons \cite{Page:1976df}. In the analysis given in Section \ref{statsys}, we consider a generic timelike particle (massless particles are assigned an effective mass via the reduced Compton wavelength $\lambda_C=\hbar/m$), and it will be shown that the generic particle can follow any distribution without affecting the statistical analysis. We then discuss in Section \ref{thermsoc} Hawking radiation in the context of thermal noise analysis, which will be addressed in a separate, follow-up report. This supposes that black holes, as off-equilibrium entities, exhibit characteristics of self-organizing criticality \cite{Bak1987}, with Hawking radiation serving as a stabilizing mechanism.

Throughout this analysis, we hope to answer the following questions:
\begin{itemize}
\item Can black holes and the Hawking radiation background be treated as a statistical system? 
\item Are the principles of statistical mechanics and quantum mechanics upheld in this analysis? 
\item More importantly, can we recover what is conventionally understood about Hawking radiation and black hole thermodynamics? 
\end{itemize}

\section{The Thermal Energy of a Black Hole} \label{statsys}

\subsection{Metric of Choice}

Particle worldlines near a black hole follow the Schwarzschild metric, written here in Eddington-Finkelstein (EF) coordinates \cite{Eddington:1924pmh, Finkelstein:1958zz} using the metric signature $(+,-,-,-)$:
\begin{equation}
ds^2=\left(dt'^2-dr^2-r^2d\Omega ^2 \right)+\frac{r_S}{r}\left(dt'\pm dr \right)^2,
\end{equation}
where $d\Omega^2=d\theta^2+\sin^2\theta d\phi^2$ represents the solid angle part of the metric. In EF coordinates, the modified time coordinate $dt'$ is defined as $dt\pm(dr^*-dr)$, where the sign depends on the particle's radial direction: either ``ingoing" ($+$) or ``outgoing" ($-$). Additionally, $r^*$ is the tortoise coordinate that satisfies $dr^*/dr=1/f(r)$ with $f(r)=(1-r_S/r)$. 

For an evaporating black hole, ingoing EF coordinates are used with the Vaidya metric instead of the Schwarzschild metric \cite{Abdolrahimi:2016emo}. The Vaidya metric, given originally in Ref. \cite{Abdolrahimi:2016emo} in the $(-,+,+,+)$ signature, is revised here for consistency with the $(+,-,-,-)$ signature:
\begin{equation}
ds^2=e^{2\psi(z)}\left(1-\frac{2G\mu(z)}{r} \right)dv^2-2e^{\psi(z)} dvdr-r^2d\Omega^2,
\end{equation}
where, defining $z=r_S/r$,
\begin{equation}
\begin{split}
&\psi(z)\simeq\frac{m_P^2}{M_0^2}\left[g(z)-4\alpha\ln(z)\right],\quad\alpha\approx3.75\times10^{-5},\\
&\mu(z)\simeq M\left[1+\frac{m_P^2}{M_0^2}h(z) \right],
\end{split}
\end{equation}
with $m_P\equiv\sqrt{\hbar/G}$ as the Planck mass, $M_0$ as the initial black hole mass (which is gauged to be astronomical), and $M$ as the instantaneous mass. We can define the ratio between the squares of the initial black hole and Planck masses as an ``initial abundance number'': 
\begin{equation} \label{number}
N_0=\frac{M_0^2}{m_P^2},
\end{equation}
such that the initial black hole's surface area $A_0=16\pi G^2M_0^2$ is divided into $N_0$ quantum black hole surface areas via $A_Q=16\pi l_P^2$ \cite{Hawking:1971ei, Carr:2005qn}, where $l_P\equiv\sqrt{G\hbar}$ is the Planck length.

The functions $g(z)$ and $h(z)$ are polynomials in $z$, given in Ref. \cite{Abdolrahimi:2016emo} as Eq. (85). If we omit higher-order terms of $z$, given the relevant range $z\in(0,1]$, we define
\begin{equation}\label{ghfull}
\begin{split}
&g(z)\simeq\beta{(1-z)(1+z)}\\
&\quad\quad\times\left[31k_5-18(\xi-k_6)+6(4f_0-k_3)\frac{(73+49z)}{(1+z)}\right],\\
&h(z)\simeq\beta{(1-z)(1+z)}\\
&\quad\quad\times\left[6\xi+55k_5+18k_6-6(4f_0-k_3)\frac{(23+35z)}{(1+z)}+\frac{144k_4}{(1+z)}\right],
\end{split}
\end{equation}
where $\beta=({2^{14}3^4 5\pi})^{-1}$, and the parameters $\xi$, $k_{3,4,5,6}$ and $f_0$ depend on particle spin. E.g. for photons (spin-1): $\xi=-1248$, $k_3=81.80$, $k_4=-770.42$, $k_5=65.38$, $k_6=-942.18$, and $f_0=2.4346$ \cite{Abdolrahimi:2016emo, Page:1976ki, Jensen:1990xbb}. Other known values of $\xi$ include $168$ for spinors (spin-1/2 particles) and $20352$ for gravitons (spin-2) \cite{Abdolrahimi:2016emo}; the $f_0$ and $k$ parameters are unknown for these cases.

To achieve a spin-independent treatment of Hawking radiation, we examine how the functions $g(z)$ and $h(z)$ impact the metric analysis. Since $\xi$ is known for all particle types, we consider only the term containing $\xi$ in both functions:
\begin{equation}\label{gh}
\begin{split}
&g(z)\approx{-18\xi\beta}(1-z)(1+z),\\
&h(z)\approx{6\xi\beta}(1-z)(1+z)=-\frac{1}{3}g(z).
\end{split}
\end{equation}
The common leading factor, $6\xi\beta$, evaluates numerically as $4.84\times10^{-4}$ for spinors, $-3.59\times10^{-4}$ for photons, and $5.86\times10^{-3}$ for gravitons. Given the small magnitude of these factors and their proximity to zero, we approximate -- however naively -- $g(z)$ and $h(z)$ to be zero. This simplifies the metric functions to the following, using $N_0$ via Eq. (\ref{number}):
\begin{equation}
\psi(z)\approx-\frac{4\alpha}{N_0}\ln(z)\quad\mathrm{and}\quad\mu(z)\approx M,
\end{equation}
which effectively recovers the original ingoing Schwarzschild metric under the assumption $e^{\psi(z)}\rightarrow1$ with a readily small $\psi(z)$ through factorization:
\begin{equation}
ds^2\simeq f(r)dv^2-2dvdr-r^2d\Omega^2.
\end{equation}
While we could have directly started with the Schwarzschild metric, it was essential to derive this approximation through the ingoing Vaidya metric, showing that the polynomial functions $g(z)$ and $h(z)$ can be neglected in a species-independent framework.

Despite proceeding with the Schwarzschild metric, we want to present metric modifications that are anticipated when taking $g(z)$ and $h(z)$ via Eq. (\ref{gh}) into proper account. Expanding out the polynomials and neglecting higher orders of $z$ due to smallness, the metric functions read as
\begin{equation}
\begin{split}
&\psi(z)\approx-\frac{2}{N_0}\left[9\xi\beta+2\alpha\ln(z)\right],\\
&\mu(z)\approx M\left(1+\frac{6\xi\beta}{N_0}\right),
\end{split}
\end{equation}
which modifies the Vaidya metric as follows, in terms of $z$:
\begin{equation}
\begin{split}
&ds^2=\mathcal{E}(z)\left[1-z\left(1+\frac{6\xi\beta}{N_0}\right) \right]dv^2\\
&\quad\quad\quad\quad\quad\quad\quad\quad-2\mathcal{E}(z)^2 dvdr-r^2d\Omega^2,\\
&\mathcal{E}(z):=\left(e^{9\xi\beta}z^{2\alpha}\right)^{N_0/4}.
\end{split}
\end{equation}
The modification's inclusion of $\xi$ suggests a more explicit treatment of spin-dependent particle types, which is encouraged to pursue for future work. While the reduced forms of the metric functions are mutual for spinors, photons, and gravitons, the analysis on photons should involve the explicit form of the polynomials via Eq. (\ref{ghfull}), as all parameters are known.

\subsection{Worldline Lagrangian}

We define the Lagrangian $\mathcal{L}$ of the emitted Hawking particle and its mirror partner by expressing the Schwarzschild metric in terms of proper-time differentials and introducing its relativistic mass $m_H$:
\begin{equation} \label{setup}
\mathcal{L}=m_H{f(r)}\dot{v}^2 -2m_H\dot{v}\dot{r}-m_Hr^2\dot\Omega^2,
\end{equation}
where we assume motion upon the equatorial plane ($\theta=\pi/2$), so that $\dot\Omega^2=\dot\phi^2$. Here, the overdot denotes differentiaton with respect to proper time, $^\bullet=d/d\tau$. For timelike particle trajectories in EF coordinates, we define $\dot{r}$ and $\dot{v}$ as follows:
\begin{equation}\label{rdot}
r(\tau)=\sqrt{r_S\tau}\implies\dot{r}=\frac{r_S}{2r}\equiv\frac{1}{2}\left(1-f(r)\right),
\end{equation}
\begin{equation}
v(\tau)=\int\frac{r(\tau)}{r(\tau)-r_S}d\tau\implies\dot{v}=\frac{1}{f(r)}.
\end{equation}
To redefine the angular term, we introduce the relation $m_Hr^2\dot\phi^2=L\dot\phi$, where the angular momentum is given by $L=m_Hr^2\dot\phi$. This choice allows us to incorportate quantum effects by recognizing that even a radially moving particle must possess an inherent angular momentum due to its energy state, which is proportional to $\hbar$. Setting $L=q\hbar$, we establish a quasi-quantum-classical correspondence:
\begin{equation}\label{qcc}
m_Hr^2\dot\phi= q\hbar.
\end{equation}
Substituting this into the Lagrangian, the expression for the Hawking particle becomes
\begin{equation}\label{lang}
\mathcal{L}=m_H-q\hbar\frac{d\phi}{d\tau}.
\end{equation}
This form reveals that the Lagrangian consists of the relativistic energy of the emitted particle and a quantum term, suggesting a discrete energy structure reminiscent of Bohr's hydrogen model. If we model black holes as structured potential wells containing $s$ total quantum states, the availability and abundance of these states depend on the black hole's size by decomposing the black hole surface into $N$ quantum areas, as initially defined as Eq. (\ref{number}) for an initial mass $M_0$ Thus, for an instantaneous mass, $N=M^2/m_P^2$. 

The emitted Hawking particle's energy state $q$ can be interpreted as an available emission state from the black hole surface. Therefore, the state $q$ has an associating emission rate $dq/d\tau$ to describe evaporation from the spherical horizon. This is paired with the rate $d\phi/d\tau$ via a total derivative on the coupled term $q\phi$, factoring in the reduced Planck constant:
\begin{equation} \label{tder}
\frac{d(q\hbar\,\phi)}{d\tau}=q\hbar\frac{d\phi}{d\tau}+\phi\hbar\frac{dq}{d\tau}.
\end{equation}
 Utilizing $q\hbar=L$ in the total derivative term (the left-hand side of Eq. [\ref{tder}]) and enforcing Eq. (\ref{qcc}) defines a torque-like quantity $d(L\phi)/d\tau$. We can therefore define $q\hbar\, d\phi/d\tau$ in Eq. (\ref{lang}) by re-arranging Eq. (\ref{tder}). This redefinition is specifically useful when considering a stationary, spherical black hole with azimuthal symmetry (i.e., $d\phi=0$ in Eq. [\ref{lang}]). This reveals that $\phi=\text{const.}$ and the Lagrangian would only have the relativistic energy term with no quantum contribution. Redefining $d\phi/d\tau$ via Eq. (\ref{tder}) not only preserves the quantum contribution via the state emission rate $dq/d\tau$, but it also introduces a torque contribution $dL/d\tau$ via Eq. (\ref{qcc}) under a constant $\phi$. Constraining $\phi=\text{const.}$ involves, effectively, integrating over all azimuthal angles to gauge $\phi=2\pi$. This yields: 
\begin{equation} \label{entot}
 \mathcal{L}=m_H+2\pi \hbar\frac{dq}{d\tau}-2\pi\frac{dL}{d\tau}.
\end{equation}

\subsection{A Generalized Liouville Theorem} \label{modliou}

In Eq. (\ref{entot}), the rate $dq/d\tau$ describes the evaporation of a quantum state from the black hole, transitioning into a corresponding quantum state in the radiation background. This process parallels Liouville's theorem, which states that in a closed Hamiltonian system, the total number of microstates $\Omega$ remains constant if the density of a 2$n$-dimesnional phase space is constant \cite{Liouville:1838zza}:
\begin{equation}\label{liou}
\frac{d\Omega}{d\tau}=\iint \frac{d^n\mathbf{x}d^n\mathbf{p}}{(2\pi\hbar)^n}\left(\partial_\tau\varrho+\vec{\nabla}\cdot\varrho\vec{v} \right)=0.
\end{equation}
Here, $\varrho$ represents the total-microstate density governed by a probability distribution. 

Eq. (\ref{liou}) is applicible to classical Hamiltonian systems, typically for $n=3$. To generalize this for relativistic Hamiltonian systems, the set of accessible microstates evolves in a way that must be consistent with the ergodic hypothesis across all possible foilations \cite{Torrieri:2023udd}. This suggests that the rate of change of $\Omega$ (here in a 8D phase space, i.e. $n=4$) should be expressed in terms of a general relativistic continuity equation, incorporating the covariant divergence of microstate flux:
\begin{equation}
\frac{d\Omega}{d\Sigma_0}\Big|_{t-t'\simeq d\tau}=\sum_{i=1}^{N_0}\iint \frac{d^4\mathbf{x}d^4\mathbf{p}}{(2\pi\hbar)^4\sqrt{-g}}\nabla_\mu(\rho_i u^\mu)=0.
\end{equation}
Here, $\nabla_\mu$ is the covariant derivative, $\rho_i$ represents the density of a single microstate (i.e., $\varrho=\sum_i\rho_i$), and $u^\mu$ is the local 4-velocity of the microstate flow. This equation enforces phase-space volume conservation under local ergodicity, assuming $\Omega$ sufficiently explores each available state over time.

To apply this specifically to black hole evaporatation, Hawking radiation leads to a net decrease in the number of surface microstates. This is given by the the conversion of the initial number $N_0$ via Eq. (\ref{number}) into the instantaneous number $N$, such that $N\in[1,N_0]$ and $N_0-N$ quantifies the number of quantum areas already radiated from the black hole surface. To account for this,  we introduce a ``system-environment ensemble," where microstates lost from the black hole surface are transferred to the radiation background: 
\begin{equation}
\nabla_\mu\left\{\left(\rho_i^\text{BH}+\rho_i^\text{rad}+\mathcal{D}\rho_i^\mathrm{BH\rightarrow rad} \right)u^\mu \right\}=0.
\end{equation}
Here, $\rho_i^\text{BH}$ and $\rho_i^\text{rad}$ denote the microstate densities associated with the black hole and radiation field, respectively, while $\mathcal{D}\rho_i^\mathrm{BH\rightarrow rad}$ captures possible dissipative effects in the microstate transfer. To explicitly define these contributions, $\mathcal{D}$ incorporates quasi-normal mode corrections and entropy differences between subsystems:
\begin{equation}
\mathcal{D}\sim e^{-\gamma\tau}\exp\left(\mathrm{I}\omega\tau\right)\frac{S_\text{rad}-S_\text{BH}}{S_\text{rad}+S_\text{BH}}
\end{equation}
($\mathrm{I}=\sqrt{-1}$). If we assume that black hole evaporation and microstate transfer are isentropic, $S_\text{rad}=S_\text{BH}$, and dissipation vanishes: $\mathcal{D}=0$. This yields, in the local limit for one microstate such that $\nabla_\mu(\rho_i u^\mu)\rightarrow d\rho_i/d\tau$ (and dropping the Latin index), a generalized Liouville theorem:
\begin{equation} \label{lioubh}
 \frac{d}{d\tau} \iint \rho_{\text{BH}} \frac{d^4\mathbf{x} d^4\mathbf{p}}{(2\pi\hbar)^4} + \frac{d}{d\tau}\iint \rho_{\text{rad}} \frac{d^4\mathbf{x} d^4\mathbf{p}}{(2\pi\hbar)^4} = 0.
\end{equation}

This ensures conservation of total phase-space volume as the black hole evaporates. Consequently, this establishes that a black hole system obeys a modified Liouville theorem, incorporating the emitted state density $\rho_\text{BH}(\sigma,\tau)$, where $\sigma$ is an arbitrary spatial dimension in the 8-dimensional phase space.

\subsection{State Evaporation}

Considering the black hole state density contribution in Eq. (\ref{lioubh}) as the state evaporation rate $dq/d\tau$, we simplify the 8-dimensional theorem by decomposing the physical and momentum-space volumes into an 6-dimensional subspace and a seperate 2D element associated with the emitted state:
\begin{equation}
\frac{dq}{d\tau}=\frac{d}{d\tau}\iint \frac{d^{3}\mathbf{x}d^{3}\mathbf{p}}{V^3_xV^3_p}\iint \rho_\text{BH}(\sigma,\tau)\frac{d\sigma dp}{2\pi\hbar}.
\end{equation}
Here, $\int d^jx=V^j_x$ and $\int d^jp=V^j_p$ define the respective $j$-dimensional spatial and momentum volumes. This separation isolates the evolution of the emitted quantum state via $\rho_\text{BH}(\sigma,\tau)$, allowing us to track the transfer of phase-space volume from the black hole to the radiation field.

For a pure system of particles, the state density function follows the statistical distribution of those particles. In our case of Hawking radiation, it is the statisitcal distribution of the emitted particles. Since we want to maintain a general approach for the Hawking particles, we consider simultaneously:
\begin{itemize}
\item Fermi-Dirac statistics -- state occupancy is limited to one per state by the Pauli exclusion principle.
\item  Bose-Einstein statistics -- state occupancy is unrestricted at any energy level.
\item Maxwell-Boltzmann statistics -- state occupancy follows classical behavior, applicible at sufficiently high temperatures.
\end{itemize}
In order to satisty all statistical conditions, we must conclude that each emitted state is occupied by one Hawking particle at the moment of emission. Therefore, the specific form of $\rho_\text{BH}(\sigma,\tau)$ will follow a thermal distribution. 

To enforce one-particle occupancy per microstate, the probability distribution must be well-defined within the intervals $\sigma\in[l,\infty)$ (where the radiation field radially extends out to infinity from a minimum cut off at $l$, the instantaneous black hole radius) and $p\in[0,m_H]$. We define the microstate density heuristically as a square-area density:
\begin{equation}\label{bhrho}
\rho_\text{BH}(\sigma,\tau)=\frac{r_S(\tau_\mathrm{evap}-\tau)}{2\sigma^2},
\end{equation}
such that it exhibits an inverse square-law for radiation intensity. Also, $\tau_\mathrm{evap}$ is the total black hole evaporation time \cite{Pickover:1996, Carroll:1996}, given by 
\begin{equation}\label{evap}
\tau_\mathrm{evap}=\frac{5120\pi G^2M_0^3}{\hbar}
\end{equation}
for the initial black hole mass.

\begin{figure} [h!]
\centering
\includegraphics[width=0.46\textwidth]{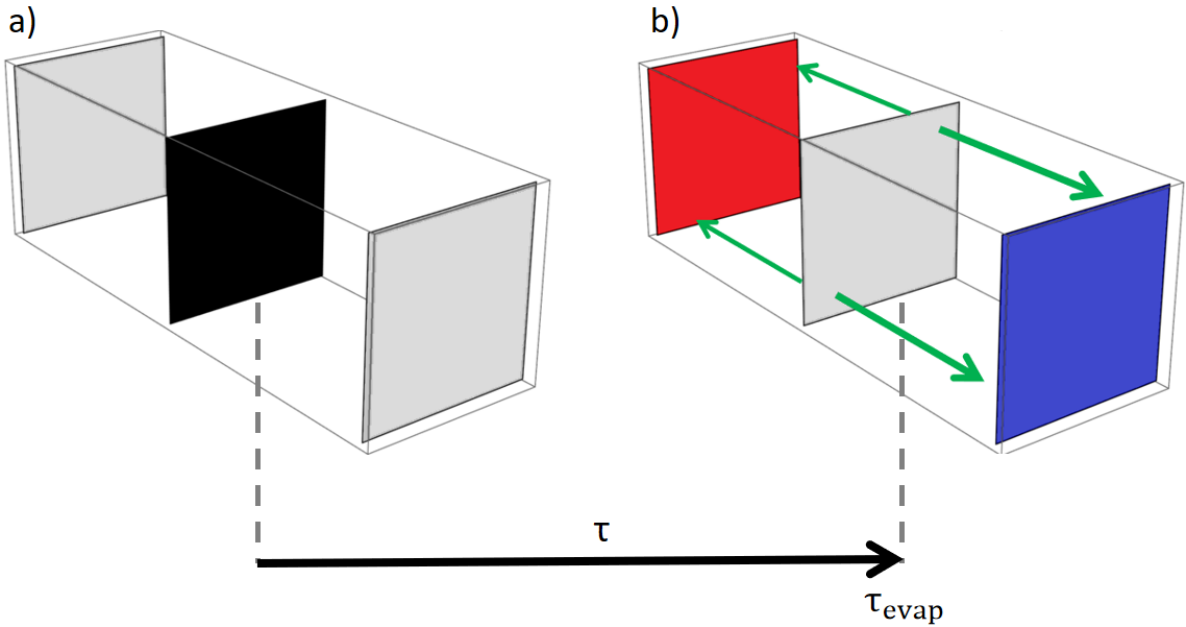}
\caption{ \label{screens} Hawking radiation of a square ``black hole cell.'' Subfigure (a) is the initial reference of a black cell between two gray screens. As time elapses toward $\tau_\text{evap}$, at which moment subfigure (b) depicts, radiation leaks from the black cell to both gray screens, counting the radiation. In subfigure (b), the ``back radiation'' is caught on the red screen, and the ``front radiation'' is caught on the blue screen. Green arrows illustrate the flow of emission states from the black cell to the radiation background, conserving the number of states within this ensemble.}
\end{figure}

Eq. (\ref{bhrho}) reflects the gradual depletion of black hole states as information transfers to the Hawking radiation field. Since Hawking quanta are emitted isotropically, the information content can be idealized as being distributed over a ``black hole cell'' with an effective square area $r_S(\tau_\mathrm{evap}-\tau)$. The quanta leaking from the horizon are recorded as bits on two additional square ``screens'' -- one in front and one behind -- each with an area of $\sigma^2$. The total information counted on these screens corresponds to the total Hawking radiation content, ensuring conservation of information in the system. A visual aid to this notion is provided in Figure \ref{screens}. 

Thus, the state evaporation rate of the black hole system is given by
\begin{equation} \label{dqdt}
\begin{split}
\frac{dq}{d\tau}&=\frac{d}{d\tau}\left((\tau_\mathrm{evap}-\tau)\frac{m_Hr_S}{4\pi\hbar} \int^\infty_l \frac{d\sigma}{\sigma^2}\right)\\
&\Rightarrow\frac{-1}{4\pi\hbar}\frac{m_H\,r_S}{l}.
\end{split}
\end{equation} 
Here, the integral's upper limit at $\sigma\rightarrow\infty$ naturally vanishes due to $1/\infty=0$, ensuring that only the boundary at $l$ contributes to the evaporation rate. 

Since the evaporation rate is, mathematically, negative and non-zero, the transfer of states from the black hole to the radiation background is a dissipative, irreversible process under semi-classical assumptions. This follows from the fact that, in this case, a quantum state escaping the horizon cannot be reabsorbed by the black hole. Upon instantaneous measurement, the black hole's instantaneous radius $l$ is equivalent to the Schwarzschild radius $r_S$. Substituting this into Eq. (\ref{dqdt}) and incorporating it into Eq. (\ref{entot}), we define the Lagrangian as a reduced relativistic term with the remaining torque contribution:
\begin{equation}
E=\frac{1}{2}m_H-2\pi\frac{dL}{d\tau}.
\end{equation}

\subsection{Hawking Mass $m_H$}

Solving for the mass of the emitted Hawking particle $m_H$ will help determine the torque contribution. To determine the mass, recall Eq. (\ref{qcc}): $m_Hr^2\dot\phi= q\hbar$. We can use our conditions of azimuthal symmetry ($\phi=2\pi$) and proper time variation in $r$ to yield
\begin{equation}\label{hqmass}
2\pi m_H\frac{d(r^2)}{d\tau}=q\hbar\implies m_H =\frac{q\hbar}{2\pi r_S},
\end{equation}
implying Eq. (\ref{rdot}). Our obtained mass depends on the energy state $q$, whose evaporation rate relies strictly on the one-particle occupancy limit. This reinforces this mass relation as the mass of one Hawking particle, and we may gauge $q=1$ as a result. Conventionally, the Hawking particle mass is obtained by the reduced Compton mass formula, $m=\hbar/\lambda_C$. Since Hawking radiation originates at the horizon, the Compton wavelength corresponds to the black hole circumference, $\lambda_C=2\pi r_S$. This requires $q=1$ in Eq. (\ref{hqmass}); refer to footnote\footnote{We eventually verify that $q=1$ is fundamental by evaluating Eq. (\ref{dqdt}) using the $q$-independent Compton mass definition of $m_H$.}. From both viewpoints, we yield
\begin{equation}\label{hawkmass}
m_H=\frac{\hbar}{4\pi GM}.
\end{equation}
Furthermore, given a quantitatively fixed $q=1$ and provided $r_Sm_H=\hbar/(2\pi)$ via Eq. (\ref{hawkmass}), $L=q\hbar$ and $L=2\pi r_Sm_H$ are both constants and equal. Therefore, either expression produces no torque when proper time-differentiated: $dL/d\tau=0$. This simplifies the energy of the Hawking particle as exactly the thermal energy of the black hole:
\begin{equation}\label{ensys}
E_\mathrm{tot}=\frac{\hbar}{8\pi GM}\equiv k_BT_H.
\end{equation}
This result is consistent with the standard Hawking temperature, reinforcing the statistical interpretation of black holes as thermodynamic systems.

\subsection{Total Energy States $s$}

Eq. (\ref{dqdt}) describes the emission of a quantum state from a black hole as its size decreases due to Hawking radiation. We now verify that $q$ indeed represents a single emitted state and that at least one of the black hole's total states is lost with each radiation event. With the Hawking particle mass defined in terms of $r_S$ and $\hbar$ via the $q$-independent Compton mass, we express the differential emission of a quantum state as
\begin{equation}
{dq}=\frac{-1}{8\pi^2}\frac{d\tau}{l}.
\end{equation}
To relate this to the black hole's evaporation, we impose the condition $d\tau=dl$, linking the change in proper time to the decrease in the black hole's instantaneous radius. Since the Hawking particle follows a worldine with $ds=d\tau$, this establishes a direct correspondence between the emitted particle's trajectory and the shrinking black hole radius. 

For the instantaneous radius, the relevant range starts at the initial radius $r_{S,0}=2GM_0$ and ends at the smallest radius achievable: $r_\text{QBH}=2l_P$. Thus, integrating over the range $l\in[r_{S,0},2l_P]$, we obtain the following, with the negative sign flipping the integration limits:
\begin{equation}
q=\frac{1}{8\pi^2}\int_{2l_P}^{r_{S,0}}\frac{dl}{l} =\frac{1}{8\pi^2}\ln\left(\frac{r_{S,0}}{2l_P} \right),
\end{equation}
or equivantly $q=\ln\left({M_0}/{m_P} \right)/{8\pi^2}$ in terms of the black hole's initial mass and the Planck mass. For black hole masses $0.000212\leq(M_0/M_\odot)<4.13\times10^{30}$ ($M_\odot=2\times10^{30}$ kg is the solar mass), we find that $\lfloor q\rfloor=1$. Here, $\lfloor \bullet\rfloor$ represents the floor function, rounding down fractional values to the nearest whole integer; this is applied on the state number, as it ensures the physical logic of whole energy states in the context of quantum systems. For masses $M>4.13\times10^{30}M_\odot\simeq8.26\times10^{60}\,\text{kg}$, we obtain $\lfloor q\rfloor>1$. To give a perspective, the mass of the observable universe is of order $10^{53}\sim10^{54}$ kg (e.g. $\sim6\times10^{22}M_\odot=1.2\times10^{53}$ kg in Ref. \cite{Gaztanaga:2023hkm}). 

For astrophysical black holes with a minimum mass of $3M_\odot$, a single quantum state is emitted each time a Hawking particle escapes the event horizon. The remaining quantum states on the black hole surface, therefore, correspond to the remaining number of surface particles along the event horizon. This defines the total number of quantum states as $\lfloor s\rfloor=N\lfloor q\rfloor$, where $N$ is given by the instantaneous mass $M$. Consequently, in terms of masses, the instantaneous total number of states is
\begin{equation} \label{totalstates}
s=\frac{M^2}{8\pi^2 m_P^2}\ln\left(\frac{M_0}{m_P} \right).
\end{equation}
Therefore, $\lfloor s_0\rfloor=N_0\lfloor q\rfloor$ is the initial total number of states.

\section{Discussion}

In this report, the statistical approach to Hawking radiation and black hole thermodynamics was explored. Referring to the questions posed in Section \ref{intro}, we demonstrated that an evaporating black hole can be treated as a statistical system, as long as the emitted horizon states are converted into radiation states, conserving the number of states in our black hole-radiation field ensemble. This respected the Liouville theorem, with the conservation of microstates in the black hole-radiation field ensemble defined as Eq. (\ref{lioubh}). Also accounting for quantum effects Hawking particle are likely subjected to, one emitted horizon state into the radiation field is occupied by one Hawking particle, no matter the species type. In this heuristic approach, we yield what is conventionally understood of Hawking radiation and black hole thermodynamics, such as the mass of a Hawking particle given as Eq. (\ref{hawkmass}), and the thermal energy of a black hole given as Eq. (\ref{ensys}).   

\subsection{Black Holes + Hawking Radiation = SOC?} \label{thermsoc}

A key consequence of treating Hawking radiation as a statistical process is the potential application to thermal noise analysis. In conventional thermal systems, particles undergo Brownian motion, but because black holes are extremely cold ($T\propto1/M$), emitted Hawking particles experience suppressed thermal motion, with fluctuations instead dominated by quantum effects.

An open question is whether black holes as statistical systems and Hawking radiation as thermal noise collectively exhibit self-organized criticality (SOC). This is a phenomenon where unstable systems self-regulate through chaotic avalanches \cite{Bak1987, Held1990}. These avalanches often follow a $1/f$ noise spectrum, where the power spectral density follows a power law, $f^{-\alpha}$, with $1<\alpha\leq2$. While black hole evaporation is gradual -- suggesting Hawking radiation does not directly trigger avalanches --, it may instead act as a slow restabilization mechanism. 

Hawking radiation has been previously modeled as thermal Gaussian noise ($\alpha=2$) \cite{Sonner:2012if}, but deviations from Gaussianity may depend on fundamental black hole parameters, potentially shifting spectral properties as the system evolves. Whether the noise spectrum transitions to avalanche-like behavior depends on the underlying black hole dynamics, e.g. anisotropies and mass intake. This can be analyzed mathematically via the Langevin \cite{lang} and Fokker-Planck \cite{fp} equations, leading to a steady-state probability distribution of the noise spectrum. Numerical approaches typically discretize the Langevin equation using a forward Euler scheme  \cite{Yuvan2021, Yuvan2022ent, Yuvan2022sym, Bier2024, MacKay:2024}. Both analytical and numerical methods require knoweldge of the potential well $V(x)$, to which the Hawking particles are dynamically bound, as well as the damping $\gamma$ and diffusion $D$ governing the fluidity of Hawking particles. The latter can be determined via AdS/CFT correspondence \cite{Kovtun:2004de}, where the fluidity of Hawking particles also depends on the shear viscosity-to-entropy density ratio $\eta/\mathcal{S}$. As stated in Section \ref{intro}, this direction will be a topic of a follow-up report.

\section*{Acknowledgments}
I thank Giorgio Torrieri for valuable discussions and suggestions regarding Section \ref{modliou}, as well as the referee's constructive comments in enhancing this work.

\end{document}